\documentclass[%
 aip,
 amsmath,amssymb,
 reprint,%
]{revtex4-1}

\usepackage{graphicx}
\usepackage{subcaption}
\usepackage{dcolumn}
\usepackage{bm}
\usepackage[normalem]{ulem}
\usepackage{comment}

\usepackage[utf8]{inputenc}
\usepackage[T1]{fontenc}
\usepackage{mathptmx}
\usepackage{etoolbox}
\usepackage{multirow}
\usepackage[normalem]{ulem}
\usepackage{xcolor}

\makeatletter
\def\@email#1#2{%
 \endgroup
 \patchcmd{\titleblock@produce}
  {\frontmatter@RRAPformat}
  {\frontmatter@RRAPformat{\produce@RRAP{*#1\href{mailto:#2}{#2}}}\frontmatter@RRAPformat}
  {}{}
}%
\makeatother
\begin{document}

\preprint{AIP/123-QED}

\title{A first-principles approach for predicting infrared optical properties of solids}
\author{Sreerag Sundaram}
\affiliation{Department of Energy Science and Engineering, Indian Institute of Technology Bombay, Powai, Mumbai - 400076. India}
 \affiliation{ 
School of Mechanical Engineering, Purdue University, West Lafayette, Indiana, USA.
}%
\author{Ziqi Guo}%
\affiliation{ 
School of Mechanical Engineering, Purdue University, West Lafayette, Indiana, USA.
}%

\author{Dudong Feng}%
\affiliation{ 
School of Mechanical Engineering, Purdue University, West Lafayette, Indiana, USA.
}%

\author{Karthik Sasihithlu}
 \email{ksasihithlu@ese.iitb.ac.in}
\affiliation{%
Department of Energy Science and Engineering, Indian Institute of Technology Bombay, Powai, Mumbai - 400076. India
}%
\author{Xiulin Ruan}%
\email{ruan@purdue.edu}
\affiliation{ 
School of Mechanical Engineering, Purdue University, West Lafayette, Indiana, USA.
}%

\date{\today}

\begin{abstract}
We present a simplified formalism for predicting infrared optical constants from first-principles calculations. 
Addressing limitations of the widely used four-parameter semi-quantum Lorentz model, the proposed approach bridges the gap between the harmonic three-parameter model and  full self-energy–based methods. By incorporating essential anharmonic effects including four-phonon scattering and phonon renormalisation, the model provides an efficient and accurate alternative while maintaining low computational cost. 
The frequency-dependent refractive indices of MgO and rutile TiO$_2$ are computed and compared with experimental data, demonstrating good quantitative agreement. The framework offers a practical approach for predicting optical properties of materials across a wide range of materials.
\end{abstract}

\maketitle


\section{Introduction}

The knowledge of wavelength-dependent optical constants is fundamental to the choice of material and subsequent design as films or coatings in a wide range of fields such as radiative cooling \cite{RC_Review_intro,ZGuo_hBN_RI,AIP_Bhrigu}, photovoltaics \cite{Intro_PV1, Intro_PV2}, heat transfer \cite{Intro_HeatTransfer}, etc. The lack of information on complete optical properties, of a particular material or in a spectral band of interest, has been a long-standing key limitation. Computational predictions offer cost and temporal advantages over experimental measurements that facilitate efficient design and development of optical systems within practical constraints. 
In this regard, first-principles methods provide a direct link between atomic-scale interactions and macroscopic optical properties. Using density functional theory (DFT), first-principles calculations have yielded reasonably accurate predictions of electronic band structures \cite{ElectronicBandStructure_DFT,ElectronicBandStructure_DFT1}. By analysing the electronic transitions, it is possible to estimate a material's wavelength-dependent optical constants \cite{RIfromBandStructure_Intro,RIfromBandStructure_Intro1,Vaspkit}. 
Accuracy can be further improved by incorporating corrections such as DFT+U \cite{DFT+U} for localized orbitals,  GW  \cite{GW} for quasi-particle excitations, and BSE  \cite{GW+BSE} for optical excitations, albeit at greater computational cost. The energies associated with these electronic excitations and subsequent electronic band transitions translate to optical behaviour in the ultraviolet, the visible, and the near-infrared (NIR) regimes of the electromagnetic spectrum. 
The mid- and far-infrared (IR) spectral ranges, which are the focus of this investigation, remain critical for characterizing a material’s thermo-radiative properties, where the optical response is predominantly governed by phonon-mediated interactions \cite{RoleOfPhonons,RoleOfPhonons1}.


The lattice dielectric function at a particular photon frequency is generally described by the Lorentz classical harmonic oscillator model (Eq.~\ref{eq:dielectric_response-Lorentz3Param}), derived from Newton's equations of motion. 
\begin{equation}
    {\epsilon (\omega)} = \epsilon_{\infty} + 
    \sum_{j} \frac{S_{j} \omega_{0,j}^2}
    {\omega_{0,j}^2 - \omega^2 - i \omega \gamma_j}  
    \label{eq:dielectric_response-Lorentz3Param}
\end{equation}
Possessing only three parameters for each phonon mode, namely, the oscillator strength ($S_j$), resonant frequency ($\omega_{0,j}$), and the damping factor ($\gamma_j$), this is also often referred to as the three-parameter model. However, this model often fails to replicate certain features in the Reststrahlen bands of heteropolar materials \cite{GervaisPiriou1974,Fugallo2018}. To address these limitations, an alternative formulation based on Maxwell's laws introduces distinct transverse optical (TO) and longitudinal optical (LO) modes frequencies, $\Omega_{j,TO}$ and $\Omega_{j,LO}$, together with corresponding damping factors, $\gamma_{j,TO}$ and $\gamma_{j,LO}$: 
\begin{equation}
    {\epsilon (\omega)} = \epsilon_{\infty} 
    \prod_{j} \frac{\Omega_{j,\rm{LO}}^2 - \omega^2 + i\gamma_{j,\rm{LO}} \omega}
    {\Omega_{j,\rm{TO}}^2 - \omega^2 + i\gamma_{j,\rm{TO}} \omega} 
    \label{eq:dielectric_response-FPSQ}
\end{equation}
%
The description of each phonon mode by four parameters results in the name, four-parameter semi-quantum (FPSQ) model. Multiple studies have used the FPSQ model to demonstrate excellent fits to experimentally obtained infrared reflectance data \cite{FPSQ1, GervaisPiriou1974,ZGuo_hBN_RI,ZhenTong_FPSQ,Stokey_FPSQ,Zherui_FPSQ}. However, we noticed that the model fails when anisotropy becomes significant or underlying symmetry is low. In this context, several noteworthy points merit attention:

(i) The inherent use of the Lyddane–Sachs–Teller (LST) relation \cite{LST} in its derivation directly implies an involvement of the LO modes. Considering the manner in which transverse electromagnetic radiation interacts with atoms in a crystal \cite{BornHuang1956,KunHuang1951}, it can be  inferred that LO modes do not directly couple with the transverse radiation. As evident from the polariton dispersion \cite{KunHuang1951}, the upper polariton branch asymptotically approaches $\omega_{\mathrm{LO}}$ as $\vec{k} \to 0$, representing the material’s longitudinal response embedded in the dielectric function rather than any direct coupling to light. 


(ii) The FPSQ model was originally developed as an empirical framework to fit experimentally obtained IR reflectance data -- based on features in the Reststrahlen band(s), to macroscopic dielectric parameters \cite{FPSQ-fitting1, GervaisPiriou1974, FPSQ-fitting2, FPSQ-fitting3}. Extending this approach to predict the optical spectra entirely from first-principles theoretical calculations is non-trivial, particularly in anisotropic materials where LO–TO splitting may be absent along certain directions. This absence leads to ambiguity in assigning LO–TO mode pairs, which is a prerequisite for the FPSQ formalism. 
(iii) Lastly, the LST relation was originally derived for a two-atom system, in which only a single longitudinal–transverse optical (LO–TO) phonon pair exists. This formulation was later extended to multi-atom systems, where multiple LO–TO mode pairs arise. However, this generalized derivation assumes certain symmetry constraints - specifically, crystals with tetragonal symmetry about each ion \cite{LST_Cochran}. As a result, for lower-symmetry crystals, the standard LST relation may not strictly hold. Further information on the region of its validity can be obtained from Refs.~\onlinecite{LST_Rosenstock,LST_THK_Barron,ShashankaMitraChapter}.
To address these limitations, we adopt a more fundamental model that systematically treats the many-body problem using thermodynamic Green's functions. First derived by Cowley \cite{Cowley1963}, the mathematical treatment allows for the sum of an infinite number of terms in the perturbation series, thereby incorporating the phonon self-energy. As a result, the method naturally accounts for mode-resolved phonon contributions and successfully captures anharmonic effects, as well as the frequency and temperature dependence of the phonon self-energy. Notably, the LO phonon frequencies do not explicitly appear as input parameters in this model  but instead emerge implicitly from the zeros of the dielectric function. Building on this framework, we adopt it as a baseline model and introduce controlled simplifications that retain the essential physics while enabling reliable predictions of the optical constants.

\section{Methodology}
\label{Sec2}
Next, we discuss modelling the interactions between EM radiation and the crystal, and more specifically, the infrared absorption of materials. Unlike Raman scattering, IR absorption involves resonant phonon creation and therefore obeys strict energy and momentum conservation conditions between the absorbed photon and the resultant phonon. The spectral lineshape of the absorption peak is governed by the phonon self-energy, which encapsulates all inelastic scattering processes. Characterising optical behaviour via calculations of phonon self-energy terms 
has been explored in detail in several studies \cite{Cowley1968, BalkanskiWallisHaro1983, Fugallo2018, Benshalom2022, Rutile_Amano}. The complexity of the self-energy expressions, as illustrated by the work of Della Valle and Procacci \cite{Self_Energy_Terms}, makes it computationally prohibitive to account for all possible interactions, even while restricting the analysis to third- or fourth-order processes. Some studies consider only a limited subset of the possible interaction pathways and demonstrate good agreement with experimental results \cite{Fugallo2018, Rutile_Amano}, although at a non-negligible computational cost. 
While such approaches are valuable for capturing inelastic and many-body effects, we instead focus on modelling  the resonant infrared response of the crystal lattice and extracting the corresponding optical constants.

\begin{figure*}
    \centering
    
    \begin{subfigure}[b]{\textwidth}
        \centering
        \includegraphics[width=0.7\linewidth]{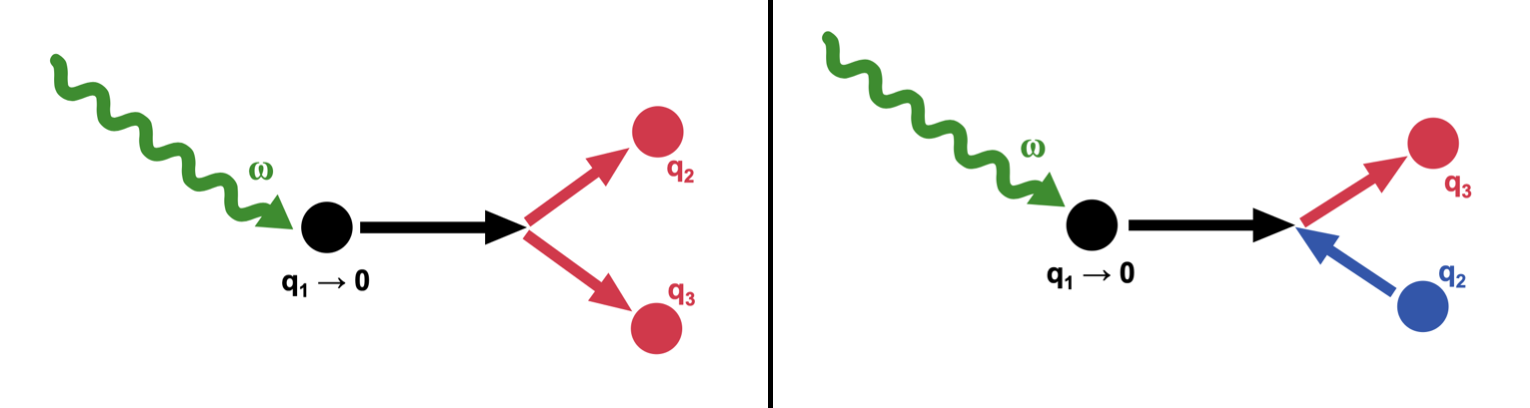}
        \caption{}\label{subfig:3Ph_interac-Pic}
    \end{subfigure}
    
    \begin{subfigure}[b]{\textwidth}
        \centering
        \includegraphics[width=\linewidth]{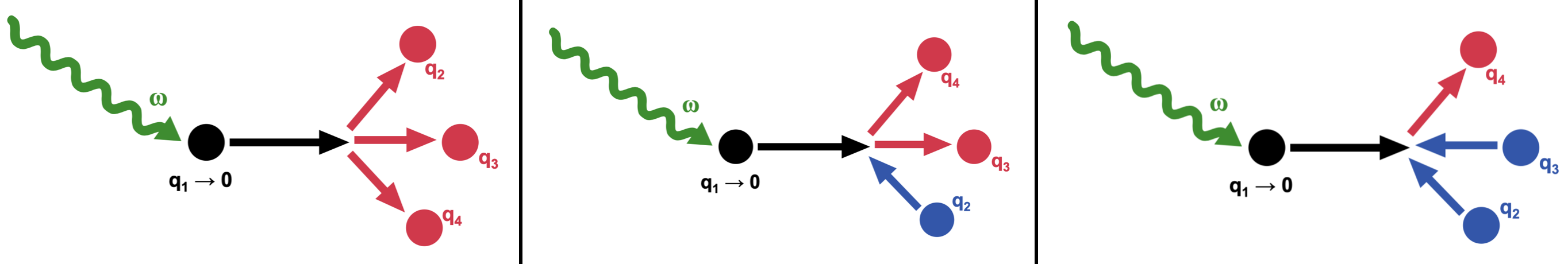}
        \caption{}\label{subfig:4Ph_interac-Pic}
    \end{subfigure}
    
    \caption{\label{fig:Interaction-Picture} Schematic of IR absorption and subsequent anharmonic phonon scattering near the Brillouin-zone centre. A photon of frequency $\omega$ excites a zone-centre optical phonon ($\mathbf{q_1} \rightarrow 0$), which undergoes (a) three-phonon and (b) four-phonon scattering processes. Energy and momentum are conserved in all processes.}
    
\end{figure*}

We model infrared absorption as a three-step process: absorption of a photon, resonant excitation of an optical phonon, and subsequent anharmonic decay of that phonon via phonon–phonon scattering, as illustrated in Fig.~\ref{fig:Interaction-Picture}. 
The resonant absorption process enforces both energy and momentum conservation between the absorbed photon and the generated phonon. Owing to the negligible momentum of infrared photons relative to Brillouin-zone dimensions, the incident photon couples predominantly to IR-active TO modes near the $\Gamma$--point ($\vec{q} \rightarrow 0$).  Absorption occurs when the photon energy matches the phonon energy, yielding a sharply resonant interaction. The generated phonon (denoted q$_1$ in Fig.~\ref{fig:Interaction-Picture}) therefore carries the same energy as the incident photon, while momentum conservation confines the excitation to wavevectors close to the zone centre.

To model the scattering rates, we follow the methodology outlined in Ref.~\onlinecite{FourPhonon}. The equations used to compute the anharmonic scattering rates for the phonon-phonon scattering, described briefly in Appendix~\ref{sec:scatt_rate}, have only been slightly altered to account for the momentum and energy conservation laws governing the interaction detailed above and depicted in Fig.~\ref{fig:Interaction-Picture}. The conservation conditions imposed significantly reduces the number of processes for which the scattering rates are to be calculated. 
An immediate effect is a substantial reduction in computational time. Statistical sampling of the complete phonon phase space \cite{Ziqi_sampling} can additionally be included for further acceleration. This also allows for quick computations of fourth order processes whose effects become pronounced for materials that are highly anharmonic.

The dielectric response is constructed through the following two-step procedure. 
In the first step,  we construct an initial self-energy corrected dielectric response as follows.   The phonon scattering rate provides the damping factor $\Gamma(\omega)$ evaluated on a sparse frequency grid corresponding to the infrared-active phonon modes near the Brillouin-zone centre. A Kramers–Kronig transformation of this sparse function yields the real part of the self-energy $\Delta(\omega)$, which represents the anharmonic shift of the phonon frequency. Using the resulting complex self-energy $\Sigma(\omega) = \Delta(\omega) - i \Gamma(\omega)$, an initial form of the dielectric function is obtained as:
\begin{equation}
\varepsilon_{\text{init},\mu\nu}(\omega,T)
=
\frac{4\pi}{\hbar  \mathrm{v}}
\sum_{0j}
M_{\mu j}M_{\nu j}
\frac{2\omega_0}
{\omega_0^2-\omega^2+2\omega_0\Sigma(\omega)} .
\label{epsilonInit}
\end{equation}
Here $ \mathrm{v} $ is the volume of the unit cell and $\omega_0$ denotes the harmonic frequency of the infrared-active TO phonon modes close to the Brillouin-zone centre. The self-energy components $\Delta(\omega)$ and $\Gamma(\omega)$ describe the anharmonic frequency shift and damping of the phonon modes respectively. M$_{{\mu}j}$ is the dipole moment along $\mu$ direction for the $j^{th}$ phonon mode. The numerator in the summation $ S^j_{\mu\nu}=M_{\mu j}M_{\nu j}2\omega_0 $, commonly referred to as the oscillator strength, is obtained from: 
\begin{equation}
\begin{aligned}
 S^j_{\mu\nu} &=
\left(\sum_{\kappa\mu'}
Z^*_{\kappa,\mu\mu'}
\frac{e_{\kappa\mu'}(\mathbf{0}j)}{\sqrt{m_\kappa}}
\right)
\left(\sum_{\kappa\nu'}
Z^*_{\kappa,\nu\nu'}
\frac{e_{\kappa\nu'}(\mathbf{0}j)}{\sqrt{m_\kappa}}
\right).
\end{aligned}
\label{eq:oscillator_strength}
\end{equation}
Here $\kappa$ denotes the atomic index, $Z^*$ is the Born effective charge tensor, $m_\kappa$ is the mass of the $\kappa$th atom, and $e_{\kappa\mu}(\mathbf{q}j)$ is the phonon eigenvector normalized such that $
\sum_{\kappa\mu} [e_{\kappa\mu}(\mathbf{q}j)]^{*} e_{\kappa\mu}(\mathbf{q}j') = \delta_{jj'}$.
Optical phonon modes exhibiting a non-zero dipole moment are infrared-active and contribute to the dielectric response, and these modes must be transversely oscillating in nature in order to couple with the electromagnetic radiation.
The poles of $\varepsilon_{\text{init}}(\omega)$ correspond to the renormalized TO phonon frequencies $\omega_j$, obtained after incorporating the anharmonic frequency shift $\Delta(\omega)$.

The expression for the dielectric function in Eq. \ref{epsilonInit} corresponds to an idealized resonant interaction in which photon absorption occurs only when the photon energy exactly matches the phonon energy. In reality, phonon–phonon interactions introduce a finite lifetime to the phonon modes, resulting in broadened absorption lines that are commonly described by a Lorentzian line shape. To account for this effect, in the second step, we reconstruct the imaginary part of the dielectric function $\varepsilon''(\omega)$ as a sum of Lorentzian contributions centered at the renormalized TO phonon frequencies obtained in Step~1, i.e.
\begin{equation}
\varepsilon''_{\mu \nu}(\omega)
=
\sum_j
A_{j, \mu \nu}
\frac{\Gamma_j^2}
{(\omega-\omega_j)^2+\Gamma_j^2},
\end{equation}
where $\omega_j$ denotes the renormalized TO phonon frequency and $\Gamma_j$ represents the phonon damping factor. The quantity $A_{j, \mu \nu}$ denotes the peak value of the imaginary part of the intermediate dielectric function $\varepsilon''_{\text{init}, \mu \nu}(\omega)$ evaluated at the corresponding renormalized TO phonon frequency. The real part of the dielectric function, $\varepsilon'_{\mu \nu}(\omega)$, is then obtained by enforcing causality through a Kramers–Kronig transformation of the reconstructed imaginary spectrum i.e., $ \varepsilon'_{\mu \nu}(\omega)  = \mathrm{KKT}\left[\varepsilon''_{\mu \nu}(\omega)\right].$
This reconstruction of $\varepsilon'_{\mu \nu}(\omega)$ requires a second Kramers–Kronig transformation to ensure that the resulting dielectric function remains causal and physically consistent. %
The resulting dielectric response $\varepsilon_{\mu \nu}(\omega)$ is obtained by combining the reconstructed dispersive component with the high-frequency contribution $\varepsilon_{\infty}$ which can be obtained from first-principles DFT calculations. The total dielectric function is therefore $\varepsilon_{\mu \nu}(\omega) =  \varepsilon_{\infty}+\varepsilon'_{\mu \nu}(\omega) + i \varepsilon''_{\mu \nu}(\omega)$, from which other relevant quantities such as the refractive index and the extinction coefficient, can be obtained.

To assess the utility of our method, we investigate the infrared optical properties of two materials, MgO and rutile TiO$_2$, using first-principles calculations at 300 K, with the aim of predicting their frequency-dependent complex dielectric function. The computed spectra show good agreement with available experimental data.  
In essence, given only the unit-cell information of a material, the proposed approach enables the extraction of frequency-dependent optical constants directly from first-principles calculations. 
To the best of our knowledge, a workflow that combines first-principles phonon scattering calculations with a self-energy–based reconstruction of the infrared dielectric response, as presented here, has not previously been reported for predicting the optical properties of materials. Importantly, the present formulation naturally incorporates higher-order phonon scattering processes, including four-phonon interactions, in determining the dielectric function.

\section{Computational Details}

DFT calculations were performed using Vienna Ab initio Simulation Package (VASP)~\cite{VASP1,VASP2,VASP3}. The Perdew-Burke-Ernzerhof (PBE) implementation of the generalized gradient approximation (GGA) \cite{GGA_PBE} was employed as the exchange-correlation function with the projector augmented wave (PAW) method \cite{VASP3}. Structural optimization was performed with electronic energy and force convergence thresholds of 10$^{-8}$~eV and 10$^{-7}$~eV/{\AA}, respectively, for both MgO and rutile TiO$_2$. Post relaxation, MgO assumes a rhombohedral  structure with  lattice parameters, $a = b = c =$ 2.975~{\AA} and $\alpha$ = $\beta$ = $\gamma$ = 60$^{\circ}$. Rutile TiO$_2$ adopts a tetragonal structure with $a = b =$ 4.662~{\AA} and c = 2.969~{\AA}. Accordingly, anisotropic behaviour is expected in rutile TiO$_2$.

Interatomic force constants (IFCs) were computed using temperature-dependent effective potentials (TDEP) \cite{TDEP,TDEP_2FC,TDEP_3FC,TDEP_4FC}, specifically the stochastic-TDEP \cite{TDEP_canonical_configuration1,TDEP_canonical_configuration2}, to capture anharmonic effects at 300~K with moderate computational cost.  Details of the method are provided in Refs.~\onlinecite{ziqi_APL,aziz_PRB} and are not repeated here. DFT calculations employed plane-wave cutoff energies of 600~eV for MgO and 500~eV for TiO$_2$, with supercells of size: 5 $\times$ 5 $\times$ 5 (MgO), and 3 $\times$ 3 $\times$ 5 (TiO$_2$). Brillouin-zone integrations were performed using  k-meshes of 3 $\times$ 3 $\times$ 3 (MgO) and 2 $\times$ 2 $\times$ 2 (TiO$_2$).  BTE calculations used q-meshes of 10 $\times$ 10 $\times$ 10 (MgO) and 6 $\times$ 6 $\times$ 10 (TiO$_2$). Harmonic IFCs were calculated including all atoms in the respective supercells.  Anharmonic IFCs were truncated using cutoff radii of 6.1~{\AA} (third order) and 5.0~{\AA} (fourth order), and  4.5~{\AA} and 4.0~{\AA} for rutile TiO$_2$.

Non-analytical term corrections were included to account for long-range dipole–dipole interactions via the Born effective charge and high-frequency dielectric tensors. Isotopic scattering was incorporated following Eq.~12 of Ref.~\onlinecite{isotopic}. 
A modified version of the FourPhonon~\cite{FourPhonon} module, built on the ShengBTE~\cite{ShengBTE_2014} framework, was used to calculate scattering rates and other parameters for the optical constants.

\section{Results and discussion}

\begin{figure*}
    \centering 

    \subfloat[\label{subfig:n-RI_MgO}]{%
        \includegraphics[page=1,width=0.45\linewidth]{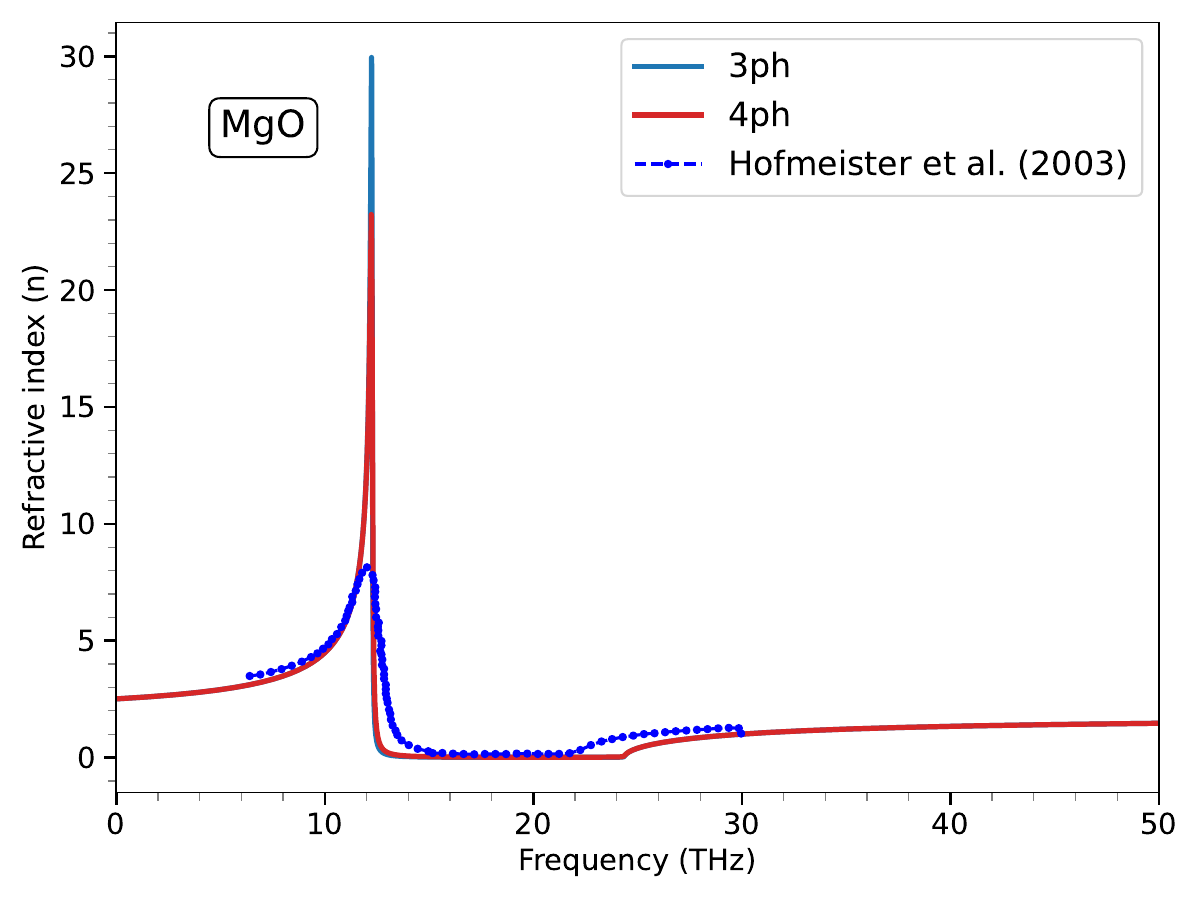}
    }
    \hfill 
    \subfloat[\label{subfig:k-RI_MgO}]{%
         \includegraphics[page=2,width=0.45\linewidth]{Images/RI-MgO-Freq.pdf}
    }

    \par\bigskip 

    \subfloat[\label{subfig:n-RI_rutile001}]{%
        \includegraphics[page=1,width=0.45\linewidth]{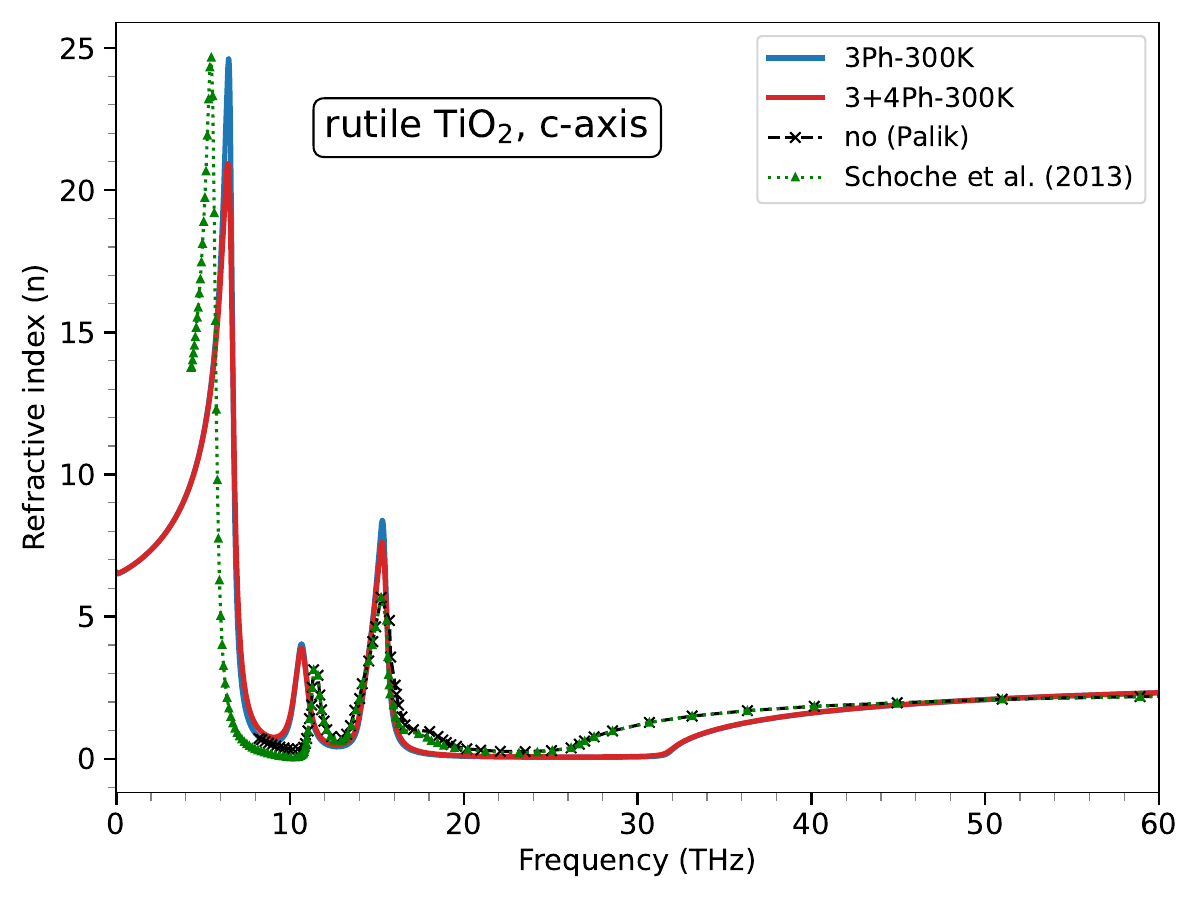}
    }
    \hfill 
    \subfloat[\label{subfig:k-RI_rutile001}]{%
         \includegraphics[page=2,width=0.45\linewidth]{Images/RI-Rutile001-oldTDEP-2.pdf}
    }

    \par\bigskip 
    
    \subfloat[\label{subfig:n-RI_rutile100}]{%
         \includegraphics[page=1,width=0.45\linewidth]{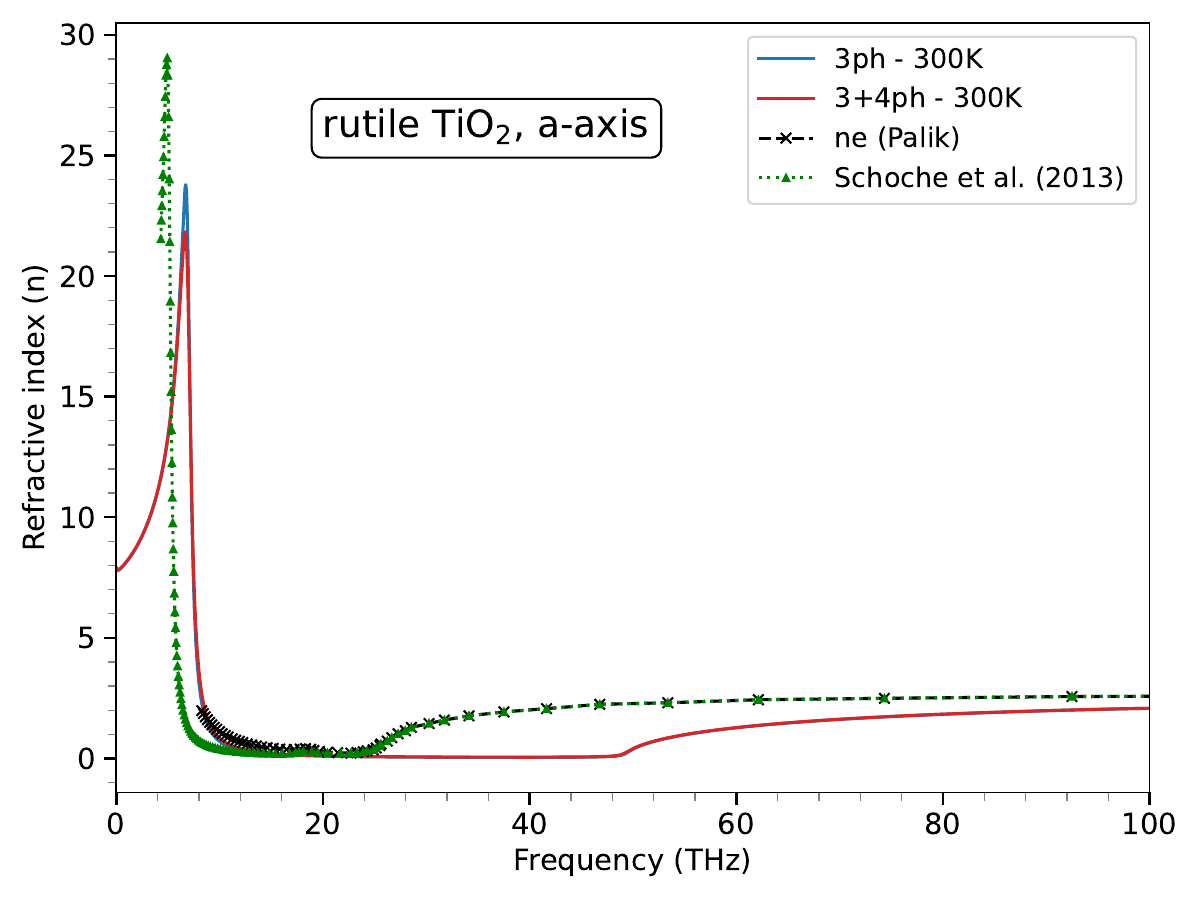}
    }
    \hfill 
    \subfloat[\label{subfig:k-RI_rutile100}]{%
         \includegraphics[page=2,width=0.45\linewidth]{Images/RI-Rutile100-Freq-2.pdf}
    }
    

    \caption{Refractive index $n$ and extinction coefficient $\kappa$ at 300~K for MgO (a,b), and rutile TiO$_2$: light propagation along the $c$-axis (ordinary response, $n_o$, $\kappa_o$) (c,d), and along the $a$-axis (extraordinary response, $n_e$, $\kappa_e$) (e,f). Solid lines denote our calculations including three-phonon (3ph) and combined three- and four-phonon (3ph+4ph) scattering, while symbols correspond to experimental data from Refs.~\onlinecite{Hofmeister_MgO_RI, TiO2_Schoche, TiO2_Palik}.}

    \label{fig:Refractive_Indices} 
\end{figure*}

\begin{figure*}
    \centering 

    \subfloat[\label{fig:Dispersion_vs_Expt_MgO}]{%
        
         \includegraphics[width=0.45\linewidth]{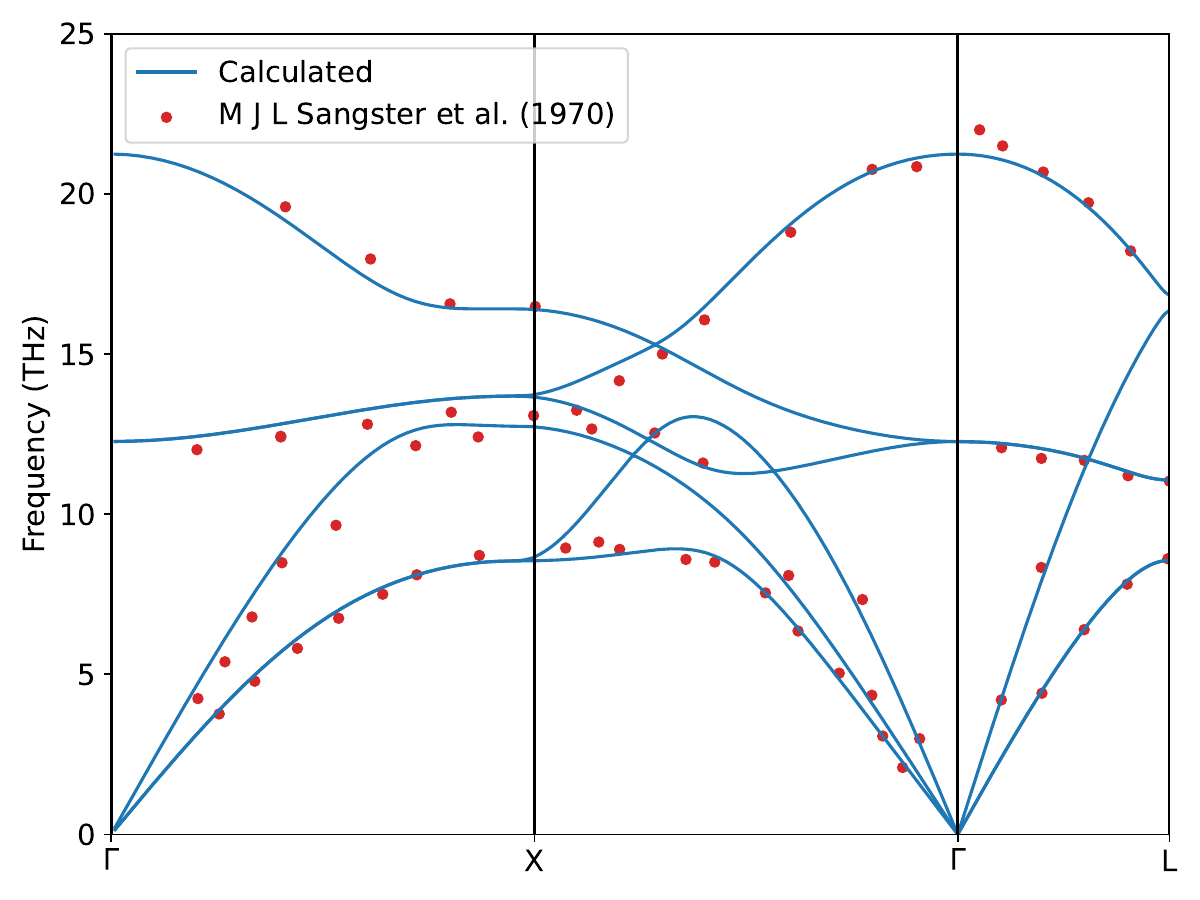}
    }
    \hfill 
    \subfloat[\label{fig:Dispersion_Rutile_vsAmano}]{%
         \includegraphics[width=0.45\linewidth]{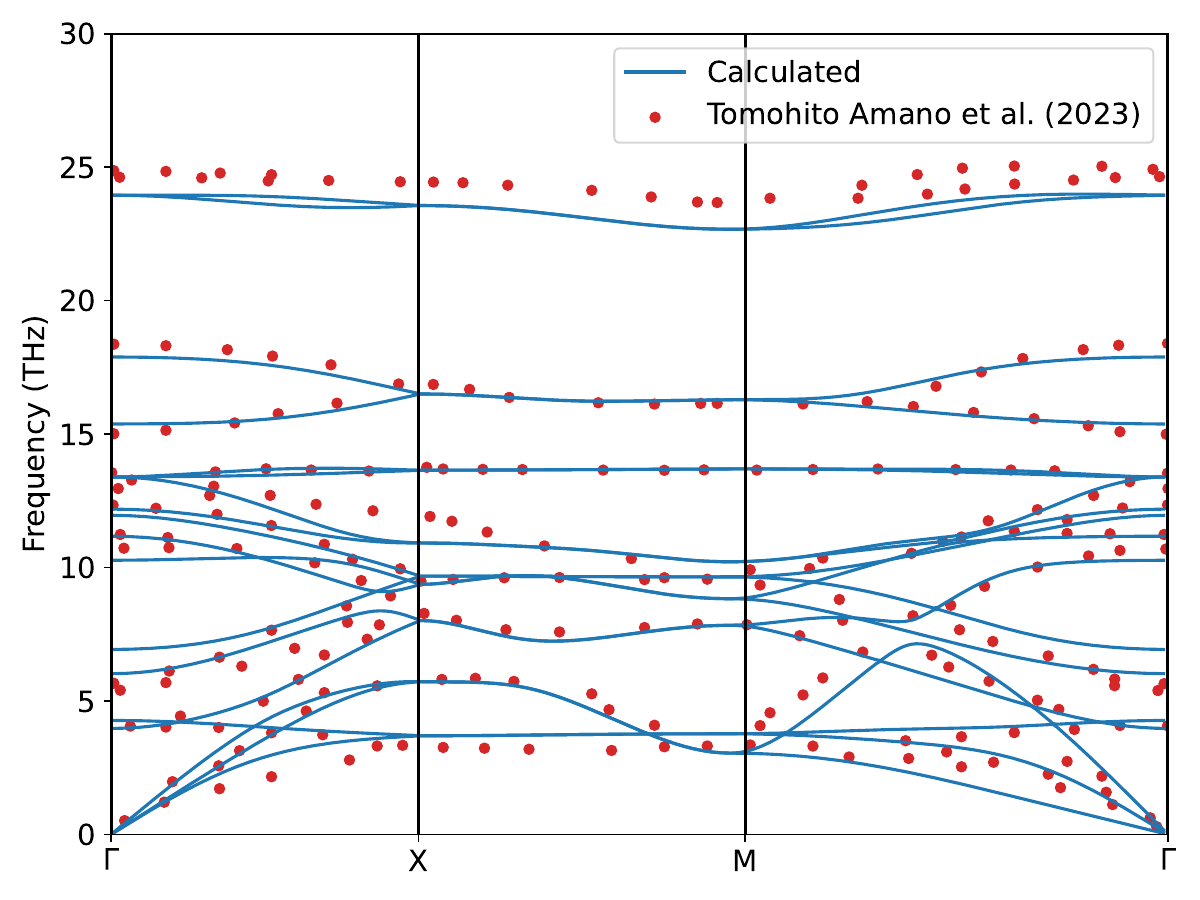}
    }
    
    \caption{Comparing the calculated phonon dispersions at 300~K using TDEP of (a) MgO with experimental data as reported in Ref.~\onlinecite{MgO_disp_expt} and (b) rutile TiO$_2$ with calculations reported in Ref.~\onlinecite{Rutile_Amano}, which themselves agree with experimental observations.}
    \label{fig:Dispersion_compare}
    
\end{figure*}

Figure~\ref{fig:Refractive_Indices} shows the calculated optical constants (complex refractive index) for MgO and rutile TiO$_2$, compared with experimental data. For MgO, the agreement with experiment  \cite{Hofmeister_MgO_RI} is good, whereas for rutile TiO$_2$ \cite{TiO2_Schoche,TiO2_Palik}, a slight shift in peak positions is observed. This discrepancy is attributed to deviations in the computed phonon dispersion, which depends sensitively on the choice of exchange–correlation functional~\cite{Rutile_Amano,GGA_LDA_TiO2_1,GGA_LDA_TiO2_2}. 
Figure~\ref{fig:Dispersion_compare} compares the phonon dispersions of MgO and rutile TiO$_2$ at 300~K with reported data from Refs.~\onlinecite{MgO_disp_expt,Rutile_Amano}, respectively. For MgO, excellent agreement is observed with inelastic neutron scattering measurements~\cite{MgO_disp_expt}. For rutile TiO$_2$, the comparison is made with calculations based on the local density approximation (LDA) exchange–correlation functional~\cite{Rutile_Amano}, in contrast to the present GGA-based results. The close agreement between the LDA dispersion and experimental data reported in Ref.~\onlinecite{Rutile_Amano} suggests that the deviations in the present results arise primarily from the choice of functional, although the overall features remain consistent with experimental observations. The shift in the optical peaks (Figs.~\ref{subfig:n-RI_rutile001} and \ref{subfig:k-RI_rutile001}) can be traced to corresponding deviations in the phonon mode frequencies. In particular, the mode associated with the dominant peak is reported at 5.15~THz~\cite{TiO2_Schoche} and 5.36~THz~\cite{Rutile_Amano}, whereas the present calculation yields a higher value of 5.99~THz at the $\Gamma$-point. This overestimation of the zone-centre phonon frequency leads directly to the observed shift in the optical spectra. The sensitivity of first-principles predictions to the choice of the exchange–correlation functional, as well as to parameters such as the high-frequency dielectric constant $\epsilon_{\infty}$, has been highlighted in Ref.~\onlinecite{Rutile_Amano}.

Recent studies by Fugallo et al.\cite{Fugallo2018} and Amano et al.\cite{Rutile_Amano} have computed  the dielectric response for MgO and rutile TiO$_2$, respectively, using the self-energy and Green's function-based approach. These formulations are based on the framework developed by Cowley~\cite{Cowley1963}, in which the interaction implicitly involves emission of a scattered photon~\cite{BalkanskiWallisHaro1983}, thereby incorporating processes beyond those required for describing resonant IR absorption. In addition, fourth-order anharmonic effects, which are non-negligible for rutile TiO$_2$ (Fig.~\ref{fig:Refractive_Indices}), are either partially treated or entirely neglected in these studies. 
The observed reduction in  peak heights and spectral broadening for rutile TiO$_2$  indicate the model's ability to capture fourth-order anharmonic effects, which are expected to become increasingly significant at elevated temperatures and in strongly anharmonic materials, as noted in Refs.~\onlinecite{Xiaolong_4Ph_ZoneCenter,ZhenTong_FPSQ}. The absence of comparable changes in MgO suggests that four-phonon interactions are negligible in this material at room temperature. 

From a computational standpoint, the approximations employed here lead to a substantial reduction in computational cost. We provide a qualitative estimate of the expected scaling  with system size. The present approach evaluates phonon-phonon scattering only for IR-active modes and restricts the calculation to  wave-vectors in the vicinity of of $\Gamma$-point. Combined with the  sampling strategy of Ref.~\onlinecite{Ziqi_sampling}, this reduces the computational complexity to approximately $O(N^2)$, where $N$ is the number of atoms in the unit cell. 
In contrast, the full self-energy formalism 
introduces an additional degree of freedom through the non-resonant frequency-dependence of the self-energy, leading to a substantially higher computational cost. We expect the scaling in such approaches to be steeper, by at least one order in $N$.

\begin{figure*}
    \centering 

    \subfloat[\label{subfig:MgO-LorentzCompare}]{%
         \includegraphics[page=2,width=0.45\linewidth]{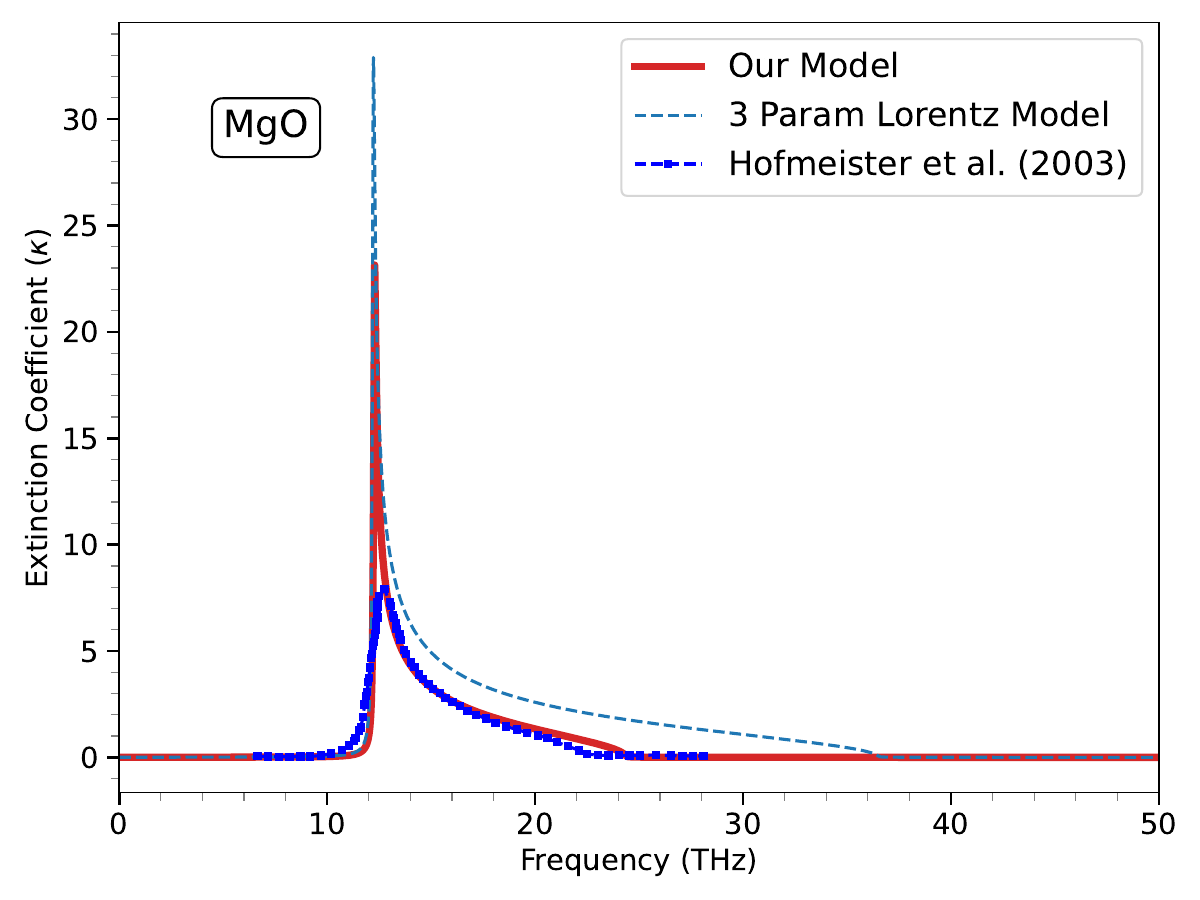}
    }
    \hfill 
    \subfloat[\label{subfig:Rutile-LorentzCompare}]{%
         \includegraphics[page=2,width=0.45\linewidth]{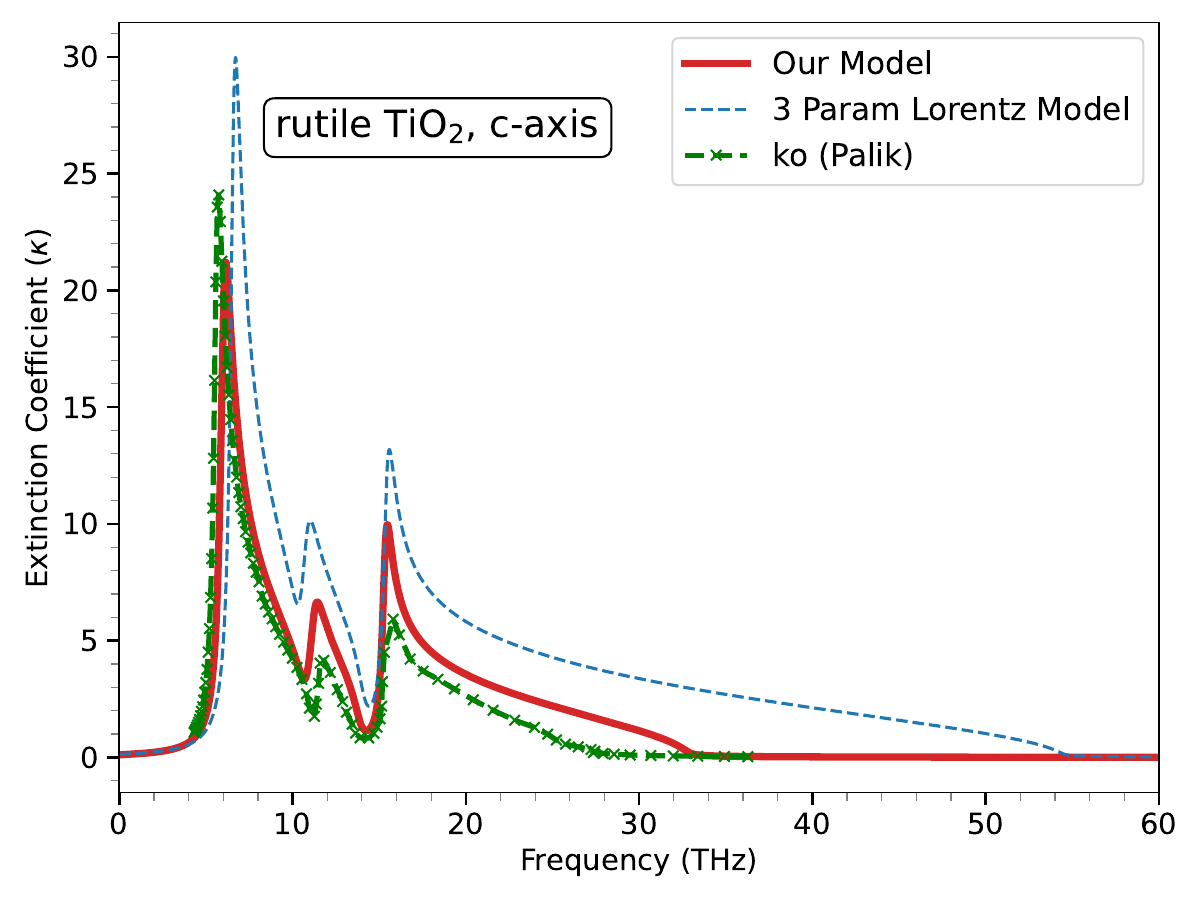}
    }
    
    \caption{Comparison of the extinction coefficient $\kappa$ calculated at 300~K using our model (red solid line) and using the three-parameter Lorentz model (blue dashed line) for (a) MgO and (b) rutile TiO$_2$. Both third- and fourth-order scattering has been included in both calculations. Experimental data reported in Refs.~\onlinecite{Hofmeister_MgO_RI, TiO2_Palik} are plotted as symbols for reference. For rutile TiO$_2$, light is incident parallel to the crystallographic c-axis, corresponding to the ordinary response (ko) in the crystal.}
    \label{fig:LorentzCompare}
\end{figure*}

Although Eq.~\ref{epsilonInit} resembles the three-parameter Lorentz model (Eq.~\ref{eq:dielectric_response-Lorentz3Param}), the two are not equivalent. In particular, the present formulation directly incorporates the anharmonic frequency shift $\Delta$. 
Figure~\ref{fig:LorentzCompare} depicts the imaginary part of the refractive index (the extinction coefficient) for MgO, and rutile TiO$_2$ along the c-axis, comparing it with predictions obtained using the three-parameter Lorentz model. The improved agreement between the model proposed in this study with experimental data, is evidence that it better captures the effects of anharmonicities. 
Furthermore, a comparison of the optical constants of rutile TiO$_2$ extracted using the presented approach and the exact self-energy formalism (Fig.~\ref{fig:Exact_SEModel_Compare}), shows close agreement indicating that the proposed model retains a high level of predictive accuracy despite its simplifications.

The scope of the present work is restricted to one-phonon processes i.e., absorption of a single photon resulting in the creation of a single phonon. Contributions from higher-order multi-phonon processes, which arise from higher-order dipole moments, are expected to be small (less than 5~\% of the total dielectric constant) \cite{Cowley1963}. Nevertheless, these contributions are expected to increase with temperature, particularly from second- and third-order dipole terms, although still small compared to the leading contribution. A more explicit treatment of such effects may therefore become relevant at elevated temperatures and represents a natural extension of the present framework.
A further extension would involve incorporating a temperature-dependent Drude contribution to the dielectric response, enabling a phenomenological description of carrier scattering in materials with metallic character, for instance arising from electron–phonon interactions. A fully microscopic treatment of these processes is expected to further improve quantitative accuracy.


\section{Conclusions}

In summary, we present a first-principles framework to compute the infrared optical properties of materials. The proposed model provides a simplified  yet physically consistent description of the interaction between infrared radiation and the crystal lattice, enabling efficient calculations while systematically incorporating quartic anharmonic effects. We discuss the limitations of the widely used four-parameter semi-quantum Lorentz (FPSQ) model and note its inadequacy for certain materials. The present formulation serves as an alternative that bridges the gap between the harmonic three-parameter Lorentz model and the full self-energy–based approach, capturing essential anharmonic effects while avoiding the high computational cost associated with the latter. The model is validated for MgO and rutile TiO$_2$ at 300~K with the extracted optical properties showing reasonable agreement with experimental data. 
The proposed approach provides a computationally efficient route for predicting the optical response of materials and is expected to be applicable across a range of systems and temperatures, with potential relevance to applications in photonics, optical coatings, and thermal management.

\appendix

\section{Anharmonic scattering rate equations} 
\label{sec:scatt_rate}
\renewcommand{\thefigure}{A.\arabic{figure}}
\setcounter{figure}{0}
\renewcommand{\thetable}{A.\Roman{table}}
\setcounter{table}{0}

The three- and four-phonon scattering rates are computed using Eqs.~\ref{eq:3ph_scatt_rate} and \ref{eq:4ph_scatt_rate} respectively. Each term retains its original physical meaning as described in Ref.~\onlinecite{FourPhonon}. Additional momentum and energy conservation constraints are applied to the phonon created from the absorbed photon (q$_1$ in Fig.~\ref{fig:Interaction-Picture}). Momentum is conserved by imposing that q$_1$ be always the point closest to the $\Gamma$--point ($q \rightarrow 0$), whereas energy conservation requires that it have the same energy as the absorbed photon, i.e. of the IR-active TO mode, also reinforced through the condition $\omega_{\lambda_1} = \omega$ in Eqs.~\ref{eq:3ph_scatt_rate} and \ref{eq:4ph_scatt_rate}. 

\begin{widetext}
    \begin{equation}
        \begin{aligned}
            \Gamma^{(+)}_{\lambda_1 \lambda_2 \lambda_3} (\omega) &= \frac{\hbar\pi}{4} \frac{n^{0}_{\lambda_2} - n^{0}_{\lambda_3}}{\omega_{\lambda_1} \omega_{\lambda_2} \omega_{\lambda_3}} |V^{(+)}_{\lambda_1 \lambda_2 \lambda_3}|^2 \delta(\omega + \omega_{\lambda_2} - \omega_{\lambda_3}) \\
            \Gamma^{(-)}_{\lambda_1 \lambda_2 \lambda_3} (\omega) &= \frac{\hbar\pi}{4} \frac{n^{0}_{\lambda_2} + n^{0}_{\lambda_3} + 1}{\omega_{\lambda_1} \omega_{\lambda_2} \omega_{\lambda_3}} |V^{(-)}_{\lambda_1 \lambda_2 \lambda_3}|^2 \delta(\omega - \omega_{\lambda_2} - \omega_{\lambda_3}) \quad \text{such that } \omega_{\lambda_1} = \omega
            \label{eq:3ph_scatt_rate}
        \end{aligned}        
    \end{equation}
    
    \begin{equation}
        \begin{aligned}
            \Gamma^{(++)}_{\lambda_1 \lambda_2 \lambda_3 \lambda_4} (\omega) &= \frac{\hbar^2\pi}{8N} \frac{(1+n^{0}_{\lambda_2}) (1+n^{0}_{\lambda_3}) n^{0}_{\lambda_4}}{n^{0}_{\lambda_1}} |V^{(++)}_{\lambda_1 \lambda_2 \lambda_3 \lambda_4}|^2 \frac{\delta(\omega + \omega_{\lambda_2} + \omega_{\lambda_3} - \omega_{\lambda_4})}{\omega_{\lambda_1} \omega_{\lambda_2} \omega_{\lambda_3} \omega_{\lambda_4}} \\
            \Gamma^{(+-)}_{\lambda_1 \lambda_2 \lambda_3 \lambda_4} (\omega) &= \frac{\hbar^2\pi}{8N} \frac{(1+n^{0}_{\lambda_2})n^{0}_{\lambda_3} n^{0}_{\lambda_4}}{n^{0}_{\lambda_1}} |V^{(+-)}_{\lambda_1 \lambda_2 \lambda_3 \lambda_4}|^2 \frac{\delta(\omega + \omega_{\lambda_2} - \omega_{\lambda_3} - \omega_{\lambda_4})}{\omega_{\lambda_1} \omega_{\lambda_2} \omega_{\lambda_3} \omega_{\lambda_4}} \\
            \Gamma^{(--)}_{\lambda_1 \lambda_2 \lambda_3 \lambda_4} (\omega) &= \frac{\hbar^2\pi}{8N} \frac{n^{0}_{\lambda_2} n^{0}_{\lambda_3} n^{0}_{\lambda_4}}{n^{0}_{\lambda_1}} |V^{(--)}_{\lambda_1 \lambda_2 \lambda_3 \lambda_4}|^2 \frac{\delta(\omega - \omega_{\lambda_2} - \omega_{\lambda_3} - \omega_{\lambda_4})}{\omega_{\lambda_1} \omega_{\lambda_2} \omega_{\lambda_3} \omega_{\lambda_4}} \quad \text{such that } \omega_{\lambda_1} = \omega
            \label{eq:4ph_scatt_rate}
        \end{aligned}        
    \end{equation}

\end{widetext}

\section{Extracting refractive index and extinction coefficient} 
\label{sec:Calc_dielectric_const+anisotropy}
\renewcommand{\thefigure}{B.\arabic{figure}}
\setcounter{figure}{0}
\renewcommand{\thetable}{B.\Roman{table}}
\setcounter{table}{0}


Dielectric response $\epsilon(\omega)$ calculated using the procedure explained in Sec. \ref{Sec2}  is substituted in Eq.~\ref{eq:n_kappa_from_eps} to obtain the refractive index $n(\omega)$ and the extinction coefficient $\kappa(\omega)$.
\begin{equation}
    \begin{aligned}
        n(\omega)^2 = \frac{1}{2} \bigg[\sqrt{\text{Re}[\epsilon(\omega)]^2 + \text{Im}[\epsilon(\omega)]^2} + \text{Re}[\epsilon(\omega)]\bigg] \\
        \kappa(\omega)^2 = \frac{1}{2} \bigg[\sqrt{\text{Re}[\epsilon(\omega)]^2 + \text{Im}[\epsilon(\omega)]^2} - \text{Re}[\epsilon(\omega)]\bigg] 
        \label{eq:n_kappa_from_eps}
    \end{aligned}
\end{equation}

\section{Comparison with the exact self-energy method} 
\label{sec:Compare_Exact_SE_Method}
\renewcommand{\thefigure}{C.\arabic{figure}}
\setcounter{figure}{0}
\renewcommand{\thetable}{C.\Roman{table}}
\setcounter{table}{0}



    

\begin{figure*}
    \centering 

    \subfloat[\label{subfig:Exact_n}]{%
         \includegraphics[page=1,width=0.45\linewidth]{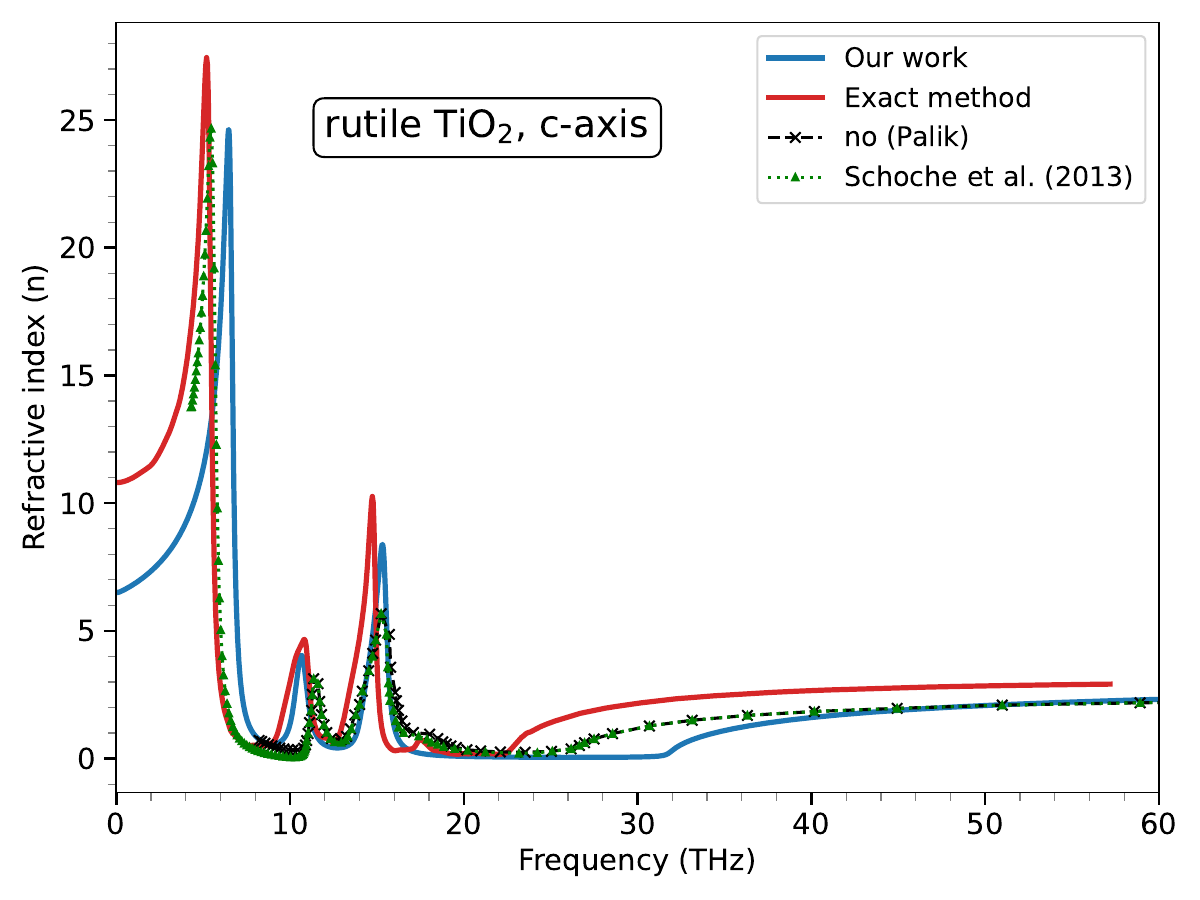}
    }
    \hfill 
    \subfloat[\label{subfig:Exact_kappa}]{%
         \includegraphics[page=2,width=0.45\linewidth]{Images/RI-ExactComp-Rutile001-2.pdf}
    }
    
    \caption{Comparison of refractive index $n$ (a) and extinction coefficient $\kappa$ (b) of rutile TiO$_2$ at 300~K, as predicted by our model (blue) and by the exact self-energy model implemented in TDEP (red). Experimental data from Refs.~\onlinecite{TiO2_Schoche, TiO2_Palik} are shown as symbols, demonstrating close agreement of both predictions. The probe light is incident along the crystal's c-axis.}
    \label{fig:Exact_SEModel_Compare}
\end{figure*}

The exact model to calculate the self-energies of phonon modes, as detailed in Refs.~\onlinecite{Cowley1968, BalkanskiWallisHaro1983, Fugallo2018, Benshalom2022, Rutile_Amano}, is implemented in TDEP \cite{TDEP_lineshape, Benshalom2022}, albeit considering three-phonon interactions only. Figure~\ref{fig:Exact_SEModel_Compare} compares the refractive index and the extinction coefficient of rutile TiO$_2$ extracted using the exact formulation and the method proposed in this study.

As discussed in the main script, the formalism of the exact model allows for more scattering processes which increases the computational cost significantly. Furthermore, the scattering rates are calculated for all phonon modes at the $\Gamma$-point, as opposed to our selective calculation involving only the IR-active TO phonon modes. The combined effect is a noteworthy reduction of computational time and resource requirement for the proposed model, without considerable loss in accuracy of predictions as demonstrated by Fig.~\ref{fig:Exact_SEModel_Compare}.

\begin{acknowledgments}
S.S. acknowledges financial support from SERB for his award [SB/S9/Z-03/2017-XXIII (2022)] under the Overseas Visiting Doctoral Fellowship (OVDF) program at Purdue University, West Lafayette, USA. S.S. also acknowledges the gracious hospitality of the Sustainable Energy and Nanoscale Transport Laboratory in the School of Mechanical Engineering at Purdue University, where this work was carried out. All simulations were performed at the Rosen Center for Advanced Computing (RCAC) of Purdue University. 

D.F., Z.G., and X.R. acknowledge partial support from the US National Science Foundation with Award \#2102645.

Z.G is partly supported by the System Fellows 2024 Doctoral Fellowship provided by the Purdue Systems Collaboratory at Purdue University.

S.S. also thanks Abdulaziz Alkandari at Purdue University, Dr. Zhen Tong at Sun Yat-Sen University and Dr. Florian Knoop at Link\"oping University for valuable discussions.
\end{acknowledgments}

\section*{Data Availability Statement}

The data that support the findings of this study are available from the corresponding author upon reasonable request.

\section*{References}

\bibliography{aipsamp}

\end{document}